\documentclass[12pt,preprint]{emulateapj}

\shorttitle{CIB Correlated Anisotropies}
\shortauthors{G. Lagache et al.}

\begin{document}


%________________________________________________________________
\title{Correlated Anisotropies in the Cosmic
Far-Infrared Background Detected by {\it MIPS/Spitzer}: Constraint on the Bias}


%________________________________________________________________
\author{G. \ Lagache\altaffilmark{1}, N. Bavouzet\altaffilmark{1}, 
N. Fernandez-Conde\altaffilmark{1}, N. Ponthieu\altaffilmark{1},
 T. Rodet\altaffilmark{2}, H. \ Dole\altaffilmark{1}, M.-A. Miville-Desch\^enes\altaffilmark{1}, 
J.-L. Puget\altaffilmark{1}
}

\altaffiltext{1} {Institut d'Astrophysique Spatiale (IAS), B\^atiment
121, F-91405 Orsay (France); Universit\'e Paris-Sud 11 and CNRS (UMR
8617);
[guilaine.lagache, nicolas.bavouzet, nestor.fernandez,
nicolas.ponthieu, herve.dole, mamd, jean-loup.puget]@ias.u-psud.fr}
\altaffiltext{2} {Laboratoire des signaux et syst\`emes (L2S),
Sup\'elec, 3 rue Joliot-Curie, 91190 Gif-sur-Yvette (France); 
Thomas.Rodet@lss.supelec.fr}



\begin{abstract}
We report the detection of correlated anisotropies in the Cosmic
Far-Infrared Background at 160 $\mu$m. We measure the power spectrum
in the {\it Spitzer}/SWIRE Lockman Hole field. It reveals unambiguously
a strong excess above cirrus and Poisson contributions, at spatial
scales between 5 and 30 arcminutes, interpreted as the signature
of infrared galaxy clustering. Using our model of infrared galaxy
evolution we derive a linear bias $b=1.74 \pm 0.16$. It is a factor 2
higher than the bias measured for the local {\it IRAS} galaxies. Our
model indicates that galaxies dominating the 160 $\mu$m correlated anisotropies
are at $z\sim 1$. This implies that infrared galaxies at high
redshifts are biased tracers of mass, unlike in the local Universe.

\end{abstract}


\keywords{infrared: galaxies --galaxies: evolution --
(cosmology:) large-scale structure of universe}


%
%________________________________________________________________
\section{Introduction}

The discovery of the Cosmic Far-Infrared Background (CIB) in 1996,
together with recent cosmological surveys from the mid-infrared to the
millimeter has revolutionized our view of star formation at high
redshifts.  It has become clear, in the last decade, that infrared
galaxies contribute to a large part of the whole galaxy build- up in the
Universe. Since the discovery of the CIB, new results on the
identification of the sources contributing to the CIB, their redshift
distribution, and their nature, are coming out at increasing speed,
especially through multi-wavelength analysis (see for a review Lagache
et al. \cite{Lag05}). Stacking analysis are also very promising to
probe the CIB source populations (e.g. Dole et al. \cite{dole06}, Wang
et al. \cite{wang06}, Dye et al. \cite{dye07}).  However, up to now,
very little information is available on the clustering of infrared
galaxies, although getting information on the clustering is essential
to understand their formation process and
to see how they relate to the other galaxy populations.\\

The first three-dimensional quantitative measurements of the
clustering strength of Ultra and Hyper Luminous Infrared Galaxies
(ULIRGs, HyLIRGs) at high redshifts ($z>1.5$) have been made by Blain
et al. \cite{blain04}, Farrah et al. \cite{farrah06} and Magliocchetti
et al. \cite{maglio07}.  These studies show that ULIRGs and HyLIRGs
are associated with the most massive dark matter halos at high
redshifts, unlike in the local Universe where the star formation is
quenched in the densest environments. CIB anisotropy observations
provide a powerful complement to direct high angular resolution
observations of individual sources.  CIB fluctuations measure, at
large angular scales, the linear clustering bias 
and, at small angular scales, the nonlinear clustering within
a dark-matter halo (Cooray \& Sheth \cite{cooray02}).  They thus probe
both the dark-matter halo mass scale and the physics governing the
formation of infrared galaxies within a halo. However, up to now
correlated anisotropies have never been firmly detected. In the 
far-infrared, detection of
anisotropies is limited to the Poisson contribution (Lagache \& Puget
\cite{Lag00}, Matsuhara et al. \cite{matsu00}, Miville-Desch\^enes et
al. \cite{mamd02})\footnote{Note however that Grossan \& Smoot
\cite{grossan} report
the detection of the clustering signature at 160 $\mu$m.}.\\
We report the detection of CIB correlated
anisotropies at 160 $\mu$m in the {\it Spitzer} SWIRE Lockman Hole
field and give a first constraint on the linear bias.
The paper is organised as follow: data are presented in
Sect. \ref{sect:data}. The power spectrum is analysed in
Sect. \ref{sect:pk}. Finally a summary and discussion are given in
Sect. \ref{sect:end}.\\ Throughout this paper we use the cosmological
parameters $h=0.71,\Omega_{\Lambda}=0.73,\Omega_{m}=0.27$. For the
dark matter linear clustering we set the normalization to
$\sigma_{8}=0.8$.
%
%________________________________________________________________
\section{The Lockman hole SWIRE field: map and power spectrum}
\label{sect:data}
SWIRE has surveyed 49 square degrees distributed over 6 fields in the
northern and southern sky (Lonsdale et al. \cite{Lonsdale04}). The
Lockman Hole is the largest field with the lowest cirrus emission. It
covers about 10 square degrees at 160 $\mu$m.

\subsection{Data reduction and map}
\begin{figure*}
\plottwo{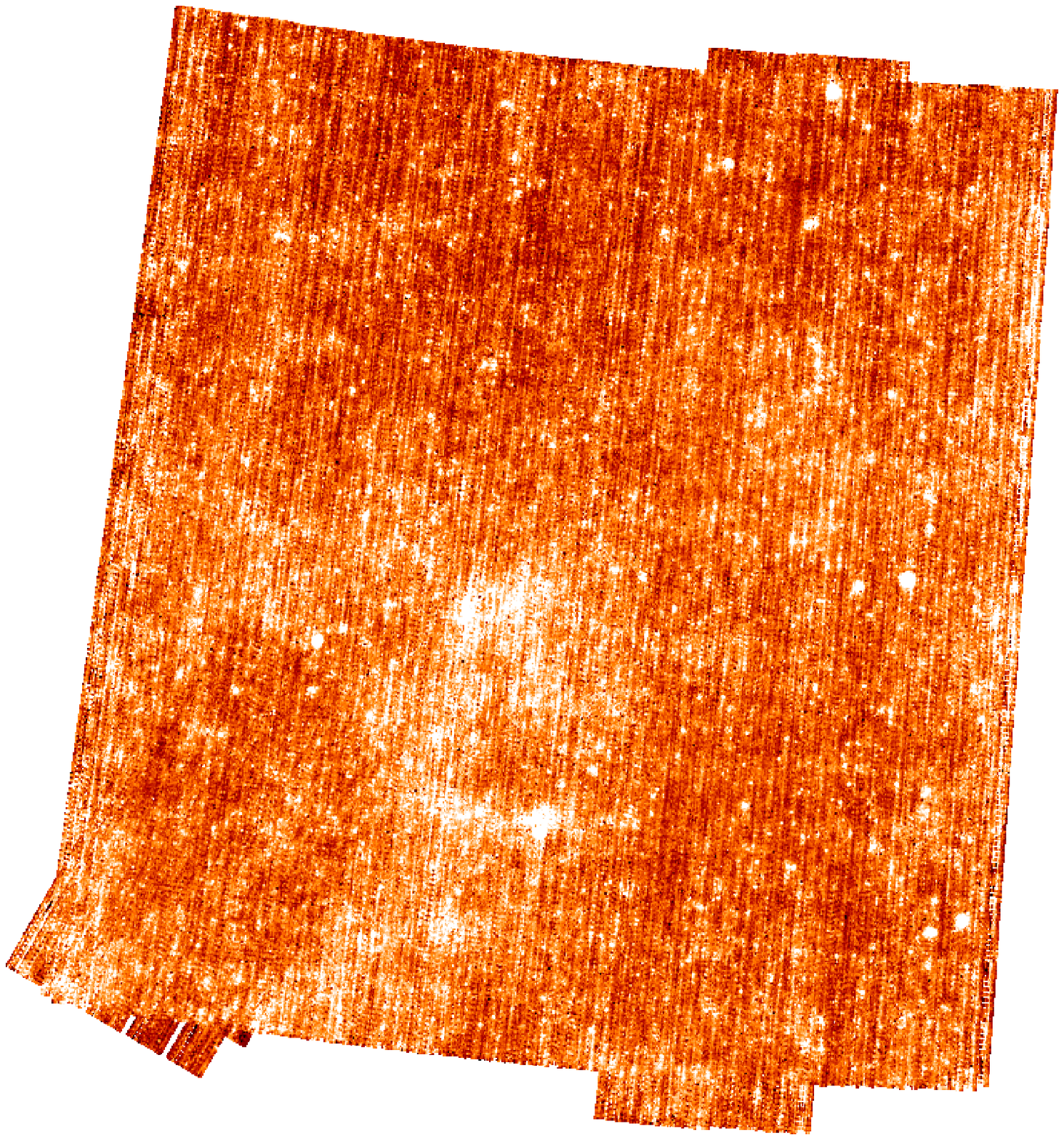}{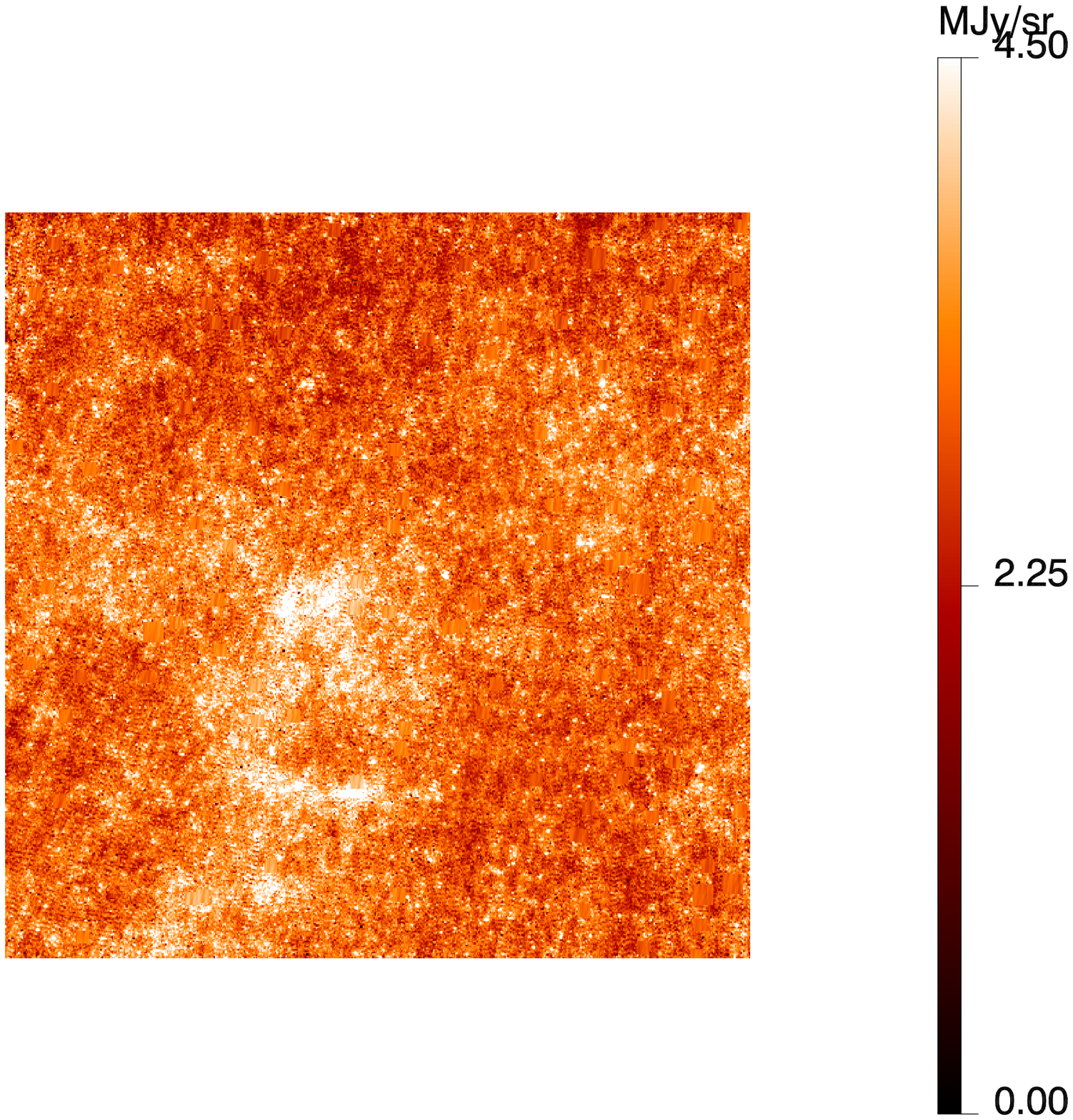}
\caption{\label{last_map} {\it Left:} 160 $\mu$m SWIRE map after standard data
reduction.  {\it Right:} Final map used to compute the power spectrum.
Residual stripes have been corrected as in Miville-Desch\^enes \&
Lagache \cite{mamd05}. Sources with $S_{160}>$200 mJy have been
removed.}
\end{figure*}

Raw data were reduced using the Data Analysis Tool (Gordon et
al. \cite{gordon05}) version 2.71. We systematically removed each DCE
(data collection event) after the stimulator flash to minimize the
latency effect. We use the last calibration factor (44.7) to convert
{\it MIPS} units to MJy/sr.  Data were finally projected on a grid with
15.95 arcsecond pixels.
The map is shown on Fig. \ref{last_map}.  Further processings were
necessary prior to measuring the power spectrum.  We first needed to
remove residual stripes. This has been done exactly in the same
way as in Miville-Desch\^ene \& Lagache
\cite{mamd05}. This paper shows that our method efficiently removed
residual stripes without affecting the astrophysical signal.  We then
remove all sources with S$_{160}>$200 mJy (200 mJy is the high
reliability threshold, as detailed in Surace et
al. \cite{surace05}). For this purpose we use DAOPHOT to detect the
sources (the image was wavelet filtered prior to the detection). We
then measure the fluxes using aperture photometry. After a fine
centering on the sources, we integrate within 25''. Sky values are
estimated in an [80''-110''] annulus. We compute the aperture
correction -- which is 2.02 -- using an effective instrumental
function (that we call PSF for Point Spread Function) measured
directly on a {\it MIPS} 160 $\mu$m map.  Using an effective PSF
rather than the PSF computed using the STinyTim
program\footnote{http://ssc.spitzer.caltech.edu/archanaly/contributed/stinytim/}
is important to take into account the survey strategy.  In the Lockman
Hole SWIRE field, the signal-to-noise ratio was not high enough to
accuratly measure the PSF. We thus use the GTO/CDFS field in which the
integration time is 6 times that in the Lockman hole SWIRE field. We
checked that our measured fluxes at 160 $\mu$m were in very good
agreement with the SWIRE DR2 catalog (better than 10$\%$ on average).
On Fig \ref{last_map} is shown the final map that will be used to
compute the power spectrum.

\subsection{Power Spectra}

There are four contributors to the power spectrum at 160 $\mu$m: 
cirrus emission, Poisson (shot) noise from discrete unresolved
sources, CIB clustering (if any), and instrumental noise.  If the
noise and the signal are not correlated, the measured power spectrum
$P(k)$ follows:
\begin{equation}
P(k) = \gamma (k) \left[P_{cirrus}(k) + P_{sources} + P_{clus}(k) \right] + N(k)
\end{equation}
where $k$ is the 2D wavenumber ($k=\sqrt{k_x^2 + k_y^2}$, expressed in arcmin$^{-1}$), 
$P_{cirrus}(k)$, $P_{sources}$, $P_{clus}(k)$ and $N(k)$ 
are respectively the power spectrum of the 
dust emission, the shot noise from unresolved sources (constant), the
clustering and the noise. 
The factor $\gamma (k)$ represents the power spectrum of the PSF.
To isolate the astrophysical components, 
we have to determine $N(k)$ and $\gamma (k)$.
\\
The noise power spectrum $N(k)$ is computed by subtracting two maps of
exactly the same region as detailed in Miville-Desch\^enes et
al. \cite{mamd02}.  We construct two maps using the even and odd
scans. As
expected, $N(k)$ and $P(k)$ meet at small scales ($k \sim 1$
arcmin$^{-1}$) where the signal is noise-dominated. The noise power
spectrum $N(k)$ is subtracted from the raw power spectrum $P(k)$.\\

\begin{figure}
\plotone{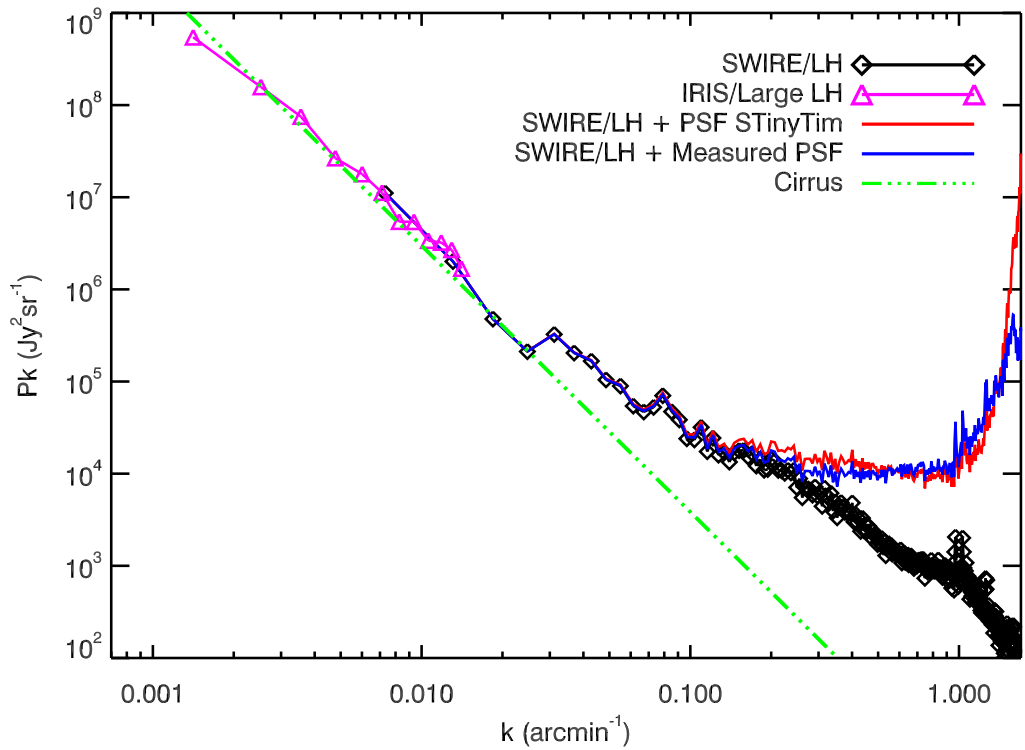}
\caption{\label{Pk_raw} Total power spectrum measured in the
Lockman Hole at 160 $\mu$m.
In black (diamonds), the {\it MIPS}/SWIRE 160 $\mu$m power spectrum
($P(k)-N(k)$, see Eq. 1), in purple (triangles) the {\it IRIS/IRAS} enlarged
Lockman Hole field power spectrum at 100 $\mu$m, scaled to the power spectrum at 160 $\mu$m using the 160/100
dust color measured at high
latitudes ($|b|>$40$^o$). In red and blue, the power spectra corrected from the
STinyTim and measured PSF -- $\frac{P(k)-N(k)}{\gamma(k)}$--,
respectively. The green dashed-3dotted line represents the best fit
cirrus power spectrum, as computed in Sect. \ref{allfit}.
}
\end{figure}

One of the critical issues is to correct the power spectrum from the
PSF $\gamma (k)$.  The PSF at 160 $\mu$m computed using the STinyTim
program is very accurate but does not include any effect induced by
the observing strategy.  We have therefore also extracted direcly the
PSF from the data (as discussed in Sect. \ref{sect:data}).  The
comparison of the power spectrum corrected by these two PSF is shown
on Fig. \ref{Pk_raw}. They are in very close agrement but we can
notice that the effective PSF gives a better result (i.e. a flat power
spectrum for $0.25<k<0.8$ arcmin$^{-1}$, as expected from
$P_{sources}$).
\\

\begin{figure*}
\plotone{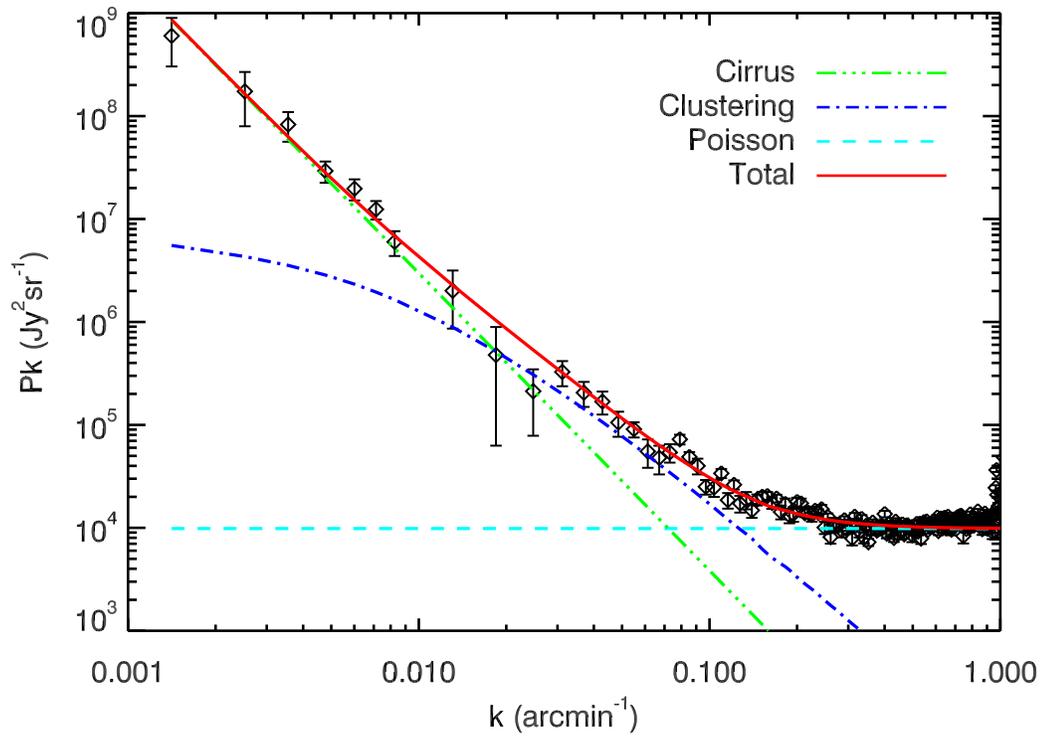}
\caption{\label{Pk_tot} Power spectrum in the Lockman Hole at 160
$\mu$m (black diamonds with
error bars) with the
three components: cirrus (green dashed-3dotted line), clustering
(blue, dashed-dotted line), Poisson (light
blue, dashed line). The red continuous line is the sum of the three components.
}
\end{figure*}

The error bars (shown in Fig. \ref{Pk_tot}) are estimated using a frequentist approach. Mock
signal plus noise maps are generated and analysed with the same
pipeline as for the data. This gives a set of power spectra of which
we compute the covariance matrix. The diagonal elements of the
covariance matrix are the errors on the measured power spectrum.

\section{Power spectrum analysis}
\label{sect:pk}

\subsection{Adding low spatial frequency data to constrain the cirrus
component} Several studies show that the cirrus component dominate the
power spectra at large scales for k$<$0.01 arcmin$^{-1}$
(e.g. Miville-Desch\^enes et al. \cite{mamd02, mamd07}). With the SWIRE
data only, in such a low interstellar dust column density field,
$P_{cirrus}$ cannot be constrained.  Larger maps are needed. We
therefore compute the power spectrum of a large ($\sim$200 Sq. Deg.) {\it IRIS/IRAS} 100
$\mu$m maps (Miville-Desch\^enes \& Lagache
\cite{mamd05}). 
The SWIRE Lockman Hole field is embedded in this large {\it IRIS} 
map so that the average 100 $\mu$m
dust emission is the same in the SWIRE and larger map (4\%
difference). Having the same average brightness is important since the
normalisation of the cirrus power spectrum in the very diffuse region
scales as $I_{100}^2$ (Miville-Desch\^enes et al. \cite{mamd07}). We
compute the power spectrum of the 100 $\mu$m map after removing the
bright sources as in Miville-Desch\^enes et al. \cite{mamd07}.  We
keep only the largest scales ($k<$9 $10^{-3}$ arcmin$^{-1}$), where we
have only the contribution from the cirrus component -- the
CIB being negligible at these very large scales -- and multiply the power
spectrum by the average dust emission color $(I_{160}/I_{100})^2$. The
color has been computed using DIRBE and FIRAS data. We compute the
average $|b|>40^o$ spectrum of the dust emission correlated with the
HI gas as in Lagache \cite{lag03}. We then fit the peak of the dust
emission spectrum to get the color. We obtain
$I_{160}/I_{100}$=2.06. If we take $|b|>30^o$, the color varies by
$\sim$10$\%$. We show on Fig. \ref{Pk_raw} the 160 $\mu$m power
spectrum derived from {\it IRIS} data together with the 160 $\mu$m
{\it MIPS} power spectrum. The spectra agree impressively well. We can
thus use this extended $P(k)$ to constrain the cirrus contribution.  In
the following, the two spectra are stiched so that we use one spectrum
from k$\sim$0.001 to 1 arcmin$^{-1}$.

\subsection{Detection of an excess at intermediate scales: signature
of correlated anisotropies}
\label{sec:excess}
The cirrus component follows:
\begin{equation}
\label{pk}
P_{cirrus}(k) = P_0 \left(\frac{k}{k_0} \right)^\beta
\end{equation}
where $P_0$ is the power spectrum value at $k_0$=0.01 arcmin$^{-1}$.
$P_0$ and $\beta$ are determined by fitting the power spectrum
(see Sect. \ref{allfit}). We obtain $P_0=(2.98 \pm 0.66) \times 10^6$
Jy$^2$/sr and $\beta =
-2.89 \pm 0.22$. The normalisation can be converted at 100 $\mu$m
using the 160/100 color given above; we obtain $P_0(100 \mu m)=7 \times 10^5$
Jy$^2$/sr.
Considering that the mean cirrus value at 100 $\mu$m in our field is
I$_{100}$=0.51 MJy/sr,
our measured $P_0(100 \mu m)$ and $\beta$ are in excellent agreement
with Miville-Desch\^enes et al. \cite{mamd07}.
%(Their Eq. 4 and 5, -2.7 and 7.e5).
The power spectrum of the cirrus component (Eq. \ref{pk}) is displayed on
Fig. \ref{Pk_raw}.
The measured power spectrum clearly has an excess of power w.r.t. the
cirrus contribution for $k>$0.3 arcmin$^{-1}$.
We interpret this strong excess as
the signature of correlated CIB anisotropies.\\

We model the correlated anisotropies following Knox et
al. \cite{knox01}\footnote{Making a much more
complex description of the correlated CIB anisotropies (as for
example adding the contribution from the clustering within the same
dark matter halo) and the bias is beyond the scope of this paper.}. 
Using the three-dimensional, linear-theory power spectrum of 
dark matter density fluctuations today, $P_{M}(k)$ the power
spectrum of CIB anisotropies can be written as:
\begin{equation}
\label{BigEq}
C_{l}^{\nu}=\int\frac{dz}{r^{2}}\frac{dr}{dz}a^{2}(z)\bar{j}^2(\nu,z)b^2
P_{M}(k)\vert_{k=l/r}G^{2}(z) \equiv P_{clus}(k)
\end{equation}
where $r$ is the comoving proper-motion distance, k the 3D wavenumber
($k=\sqrt{k_x^2 + k_y^2 + k_z^2}$, in Mpc$^{-1}$), a(z) the scale
factor, $\bar{j}(\nu,z)$ is the
mean infrared galaxy emissivity per unit of comoving volume,
and $G(z)$ is the linear theory growth function. $\ell$ is the angular
multipole sets using the Limber approximation, $k= \ell /r$.
We assume that the fluctuations in emissivity $\delta j/\bar{j}$
are a biased tracer of those in the mass and introduce the bias
parameter $b$, that we assume independent of redshift and scale: 
\begin{equation}
\frac{\delta j(\bold{k}, \nu, z)}{\bar{j}(\nu, z)} = b \times
\frac{\delta \rho(\bold{k}, z)}{\bar{\rho}(z)} 
\end{equation}
where $\rho$ is the dark matter density field.
We compute the emissivity using the infrared galaxy evolution model of Lagache et al. \cite{Lag04}.
This model, valid in the range 3-1000 $\mu$m, is in very
good agreement with mid-IR to far-IR
number counts, CIB observations, resolved sources redshift
distributions and local luminosity
functions and their evolution up to $z \sim 2$ (e.g. Lagache et al. \cite{Lag04},
Caputi et al. \cite{caputi06}, Dole et al. \cite{dole06}, Frayer et
al. \cite{frayer06}, Caputi et al. \cite{caputi07}).
Fixing the cosmology, the only unkown parameter in
Eq. \ref{BigEq} is the bias $b$.

\subsection{Measuring the bias}
\label{allfit}
We fit simultaneously $P_0$, $\beta$, $b$ and $P_{sources}$
using the non-linear least-squares curve fitting mpfit 
program\footnote{http://cow.physics.wisc.edu/$\sim$craigm/idl/idl.html}.
We obtain $P_0= (2.98\pm 0.66)\times 10^6$ Jy$^2$/sr, $\beta= -2.89 \pm 0.22$, $b=1.74
\pm 0.16$, and $P_{sources}= 9848 \pm 120$ Jy$^2$/sr. $P_0$ and $\beta$ have
been discussed in Sect. \ref{sec:excess}. $P_{sources}$ agrees quite
well with previous {\it ISOPHOT} determination at roughly the same
$S_160$ threshold
(Matsuhara et al. \cite{matsu00}). The fact that $P_0$, $\beta$, and
$P_{sources}$ are in very good agreement with previous measurements
give us confidence in our measurement of the linear bias $b \sim
1.7$. It is well known that in the local Universe infrared galaxies
are not biased tracers of the mass. For example, Saunders et
al. \cite{saunders92} found $b \sigma _8$=0.69$\pm$0.09 for {\it IRAS}
galaxies. Assuming $\sigma _8$=0.8 gives $b=$ 0.86. This bias is
roughly comparable to the bias of the SDSS galaxies at $z \sim 0.1$
($b \sim 1.1$, Tegmark et al. \cite{tegmark04}).  We measure an average
bias about 2 times higher in the CIB anisotropies. Figure \ref{fluc}
shows the predicted redshift contribution to the correlated
anisotropies. At $k$=0.05 arcmin$^{-1}$, anisotropies from $0.7<z<1.5$
infrared galaxies contribute for
more $\sim$65\%. Lower redshift galaxies contribute for less than
5\%. This shows that
infrared galaxies at z$\sim$1 are much more biased ($\sim$2 times) than locally.

\section{Summary and Discussion}
\label{sect:end}
\begin{figure}
\plotone{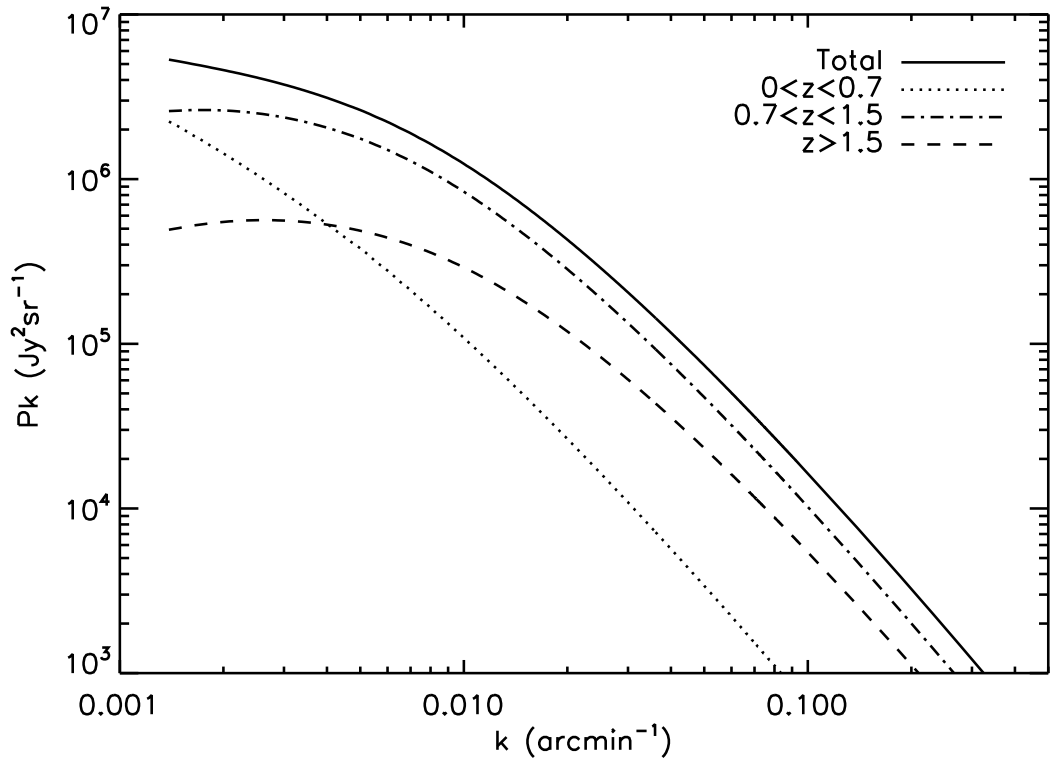}
\caption{\label{fluc} Redshift contribution to the correlated
anisotropies at 160 $\mu$m for $b=1.7$.
}
\end{figure}

We presented the power spectrum measured in the {\it Spitzer}/SWIRE Lockman Hole
field at 160 $\mu$m. It is very well reproduced by the contribution
from three components, cirrus, correlated CIB, and Poisson noise.
The cirrus and Poisson contributions are very close to previous measurements. 
We measure the linear bias, $b=1.74 \pm 0.16$.
This bias is likely to be that of infrared galaxies at $z \sim 1$
since $z \sim 1$ galaxies are dominating
the contribution to the correlated CIB anisotropies (Fig. \ref{fluc})
and local galaxies are ``anti-biased''. 
Such a bias is analog (but somewhat higher) to the bias of the red optical galaxy population at $z \sim 1$.
For example, Marinoni et al. \cite{marinoni05} measured a
bias of 1.6 at $z \sim 1.2$ for the red $(B-I) >
1.5$ galaxies in the VVDS. Blue galaxies at those redshifts are less biased
with a relative bias between red and blue population of 1.4.\\

The very strong evolution of the bias (from $\sim$1.7 at $z \sim 1 $
to 0.86 at z=0) shows that as time progresses 
and the density field evolves, nonlinear peaks become less rare events
and galaxy formation moves to lower-sigma
peaks. Thus galaxies become less biased tracers of the mass density field.
Moreover, for the infrared galaxy population it is likely that an additional
mecanism contributes significantly to the ``debiasing'' at low redshift.
Galaxies in dense environments are found to have suppressed star
formation rates (thus no or low infrared emission) and early morphological types
compared with those in the field.  
Environmental effects in particular are important in quenching the star formation
through gas stripping (e.g. Postman et al. \cite{postman05}), though on some
cluster outskirts, some star formation goes on (Duc et al. \cite{duc02}, Coia et al. \cite{coia05}).
The high bias found for infrared galaxies at $z \sim 1$ shows that star
formation rates of galaxies are increasing with the environment (as
also shown by e.g. Elbaz et al. \cite{elbaz07}). 
The ``merger bias'' is an alternative,
but somehow physically linked, way of boosting the bias at high
redshift. Clustering of objects that have undergone recent mergers
can be enhanced relative to the clustering of individual halos of
comparable masses (e.g.  Furlanetto \& Kamionkowski \cite{furna06}; Wetzel et
al. \cite{wetzel07}). It is known that the star formation in infrared galaxies is triggered
to some extent by mergers in dense environments (at z$\sim$1,
30-50\% of luminous infrared galaxies are major mergers). 
Thus, the high measured bias may also point to ``merger bias''.

%______________________________________________________________
\acknowledgments
This work is based on observations made with the {\it Spitzer} Space
Telescope, which is operated by the Jet Propulsion Laboratory,
California Institute of Technology under NASA contract 1407.
This work benefited from funding from the CNES (Centre National
d'Etudes Spatiales) and the PNC (Programme National de Cosmologie).
We warmly thanks Asantha Cooray for helpful comments and suggestions.
%______________________________________________________________
% FINAL REFERENCES

\clearpage

\end{document}